\documentclass[12pt]{article}
\usepackage{sw20lart}

\input tcilatex
\QQQ{Language}{
American English
}

\begin{document}

\textbf{TOWARDS MODELING OF CONSCIOUSNESS}

(Submitted to Phys. Rev. Lett., September 1, 2003)\vspace{0.5in}

E. A. Novikov\vspace{0.5in}

Institute for Nonlinear Science, University of California - San Diego, La
Jolla, California 92093-0402\vspace{0.5in}

An approach to nonlinear dynamical modeling of interaction between conscious
and automatic processes in the brain is proposed. Illustration of this
approach on the nonlinear equation for the current density in the cortex is
presented. Nonlinearity is determined by the sigmoidal firing rate of
neurons. The current density is considered as complex field with real and
imaginary parts representing automatic and conscious processes. The
interaction is due to the nonlinearity of the system. From this approach it
follows that the nonlinear dynamics for pure imaginary process (a dream) is
quite different from the dynamics of automatic process.\vspace{1.5in}

Ubiquitous phenomena are often the most difficult to explain. During the
last two centuries the problem of consciousness ($C$) was considered of
limits for any scientific explanation. However, recently a number of
prominent scientists took this problem seriously [1-3]. There are regular
conferences on $C$, which attract hundreds of neuroscientists, physicists,
mathematicians and philosophers. The possible advantages of developing a
theory of $C$ are enormous: from medicine to sophisticated robots to
pacification of religious conflicts.

From the physical-mathematical point of view it is desired to catch $C$ into
some sort of equations. It is common knowledge that $C$ is somehow connected
with the electrochemical activity in the brain. So, it seems logical to
start with equations for these processes. The brain activity revealed the
regime of self-similarity [4], which is typical for systems with strong
interaction of many degrees of freedom. Corresponding equations can be
formally written and are very complicated. However, at this stage the
precise form of the equations is not critical. One can use various
simplified models. The important question is how to connect these equations
with $C$ ?

The $C$-processes are subjective and, as far as we know, they can not be
measured directly by the objective methods, which are used for measuring
electrochemical (automatic) processes. At the same time, there are reasons
to believe that $C$-processes can interact with the automatic ($A$)
processes. We need equations for $A$-fields and $C$-fields, which interact
despite the fact that $C$-fields have a different nature and can not be
measured by the same methods as $A$-fields. In this Letter we suggest that
processes in the brain can be described by generalized ($G$) complex fields,
which have real $A$-component and imaginary $C$-component. These components
interact due to the nonlinearity of the equations. Such approach seems to be
the simplest. If necessary, more general constructions can be considered in
the future and they include hypercomplex components (quaternions),
additional space (branes [5]), additional time [6] and distributed sources
[7].

As first approximation, we assume that equations for the $G$-fields have the
same form as equations for the $A$-fields. Let us illustrate this with a
simple example. Consider model equation for the average (spatially uniform)
current density $\alpha (t)$ perpendicular to the cortical surface:

\begin{equation}
\frac{\partial \alpha }{\partial t}+k\alpha =f(\alpha +\sigma )  \tag{1}
\end{equation}
Here $k$ is the relaxation coefficient, $\sigma (t)$ is the average sensory
input and $f$ represents the sigmoidal firing rate of neurons (for example, $%
f(\alpha )=\tanh (\alpha )$). For the case of spatially nonuniform $\alpha
(t,\mathbf{x})$ and $\sigma (t,\mathbf{x})$ we can use a more general
equation, which include typical propagation velocity of signals $v$. Time
differentiation of (1), simple manipulation and addition a term with the
two-dimensional spatial Laplacian $\Delta $ gives:

\begin{equation}
\frac{\partial ^2\alpha }{\partial t^2}+(k+m)\frac{\partial \alpha }{%
\partial t}+(km-v^2\Delta )\alpha =(m+\frac \partial {\partial t})f(\alpha
+\sigma )  \tag{2}
\end{equation}
where $m$ is an arbitrary parameter. This type of equations are used for
interpretation of EEG and MEG spatial patterns (see recent paper [8] and
references therein). In this context we have parameters : $k\sim m\sim v/l$,
where $l$ is the connectivity scale.

Returning to (1), we now introduce the $G$-field $g=\alpha +i\psi $, where $%
\psi $ represents the $C$-effect. Substitution of $g$ instead of $\alpha $
into (1) gives system of two equations:

\begin{equation}
\frac{\partial \alpha }{\partial t}+k\alpha =Re\{f(\alpha +i\psi +\sigma )\}
\tag{3}
\end{equation}

\begin{equation}
\frac{\partial \psi }{\partial t}+k\psi =Im\{f(\alpha +i\psi +\sigma )\} 
\tag{4}
\end{equation}
These equations are coupled because $f(\alpha )$ is nonlinear. Thus, we got
simple scheme for the $A-C$ interaction. The same scheme can be applied to
equation (2) and to any nonlinear model equation. Note, that so-called
extra-sensory effects (if they exist) can be included in this approach by
assuming that $\sigma $ has an imaginary part. When $\alpha $ and $\sigma $
are relatively small (in a dream), asymptotically (4) gives closed equation
for $\psi $. The nonlinear term $Im\{f(i\psi )\}$ is quite different from
the nonlinear term $f(\alpha )$ for the $A$-process. For example $f(\alpha
)=\tanh (\alpha )$ gives $Im\{f(i\psi )\}=\tan (\psi )$. This may explain
why dreams have bizarre dynamics, not only in content, but also in intensity.

A lot of questions can be asked about the proposed approach. For instance,
how $C$-fields are related to our emotions, thoughts, images etc.? Note,
that we are modeling only an ``active'' part of $C$-field, which interact
with certain $A$-field. It is important to design experiments in which a
part of $C$-field is affected without substantial direct effect on
corresponding $A$-field. $C$-modeling, which is related to electromagnetic
activity in the brain, can be tested by using multi-channel EEG and MEG.
More general $C$-modeling can include connection between electromagnetic
activity and blood flow in the brain. In this case, brain imaging
(particularly, MRI) can be used. There are some technical issues involved in
such testing. But it seems doable.

Another important aspect of $C$-modeling is its connection with the first
principles. Introduction of imaginary fields (along with real fields)
produces generalizations, which potentially can be used for treating some
``ghosts'' [5,9] in theoretical physics. Experimentally, it is intriguing to
search for materials (besides the brain tissue), which can support an
imaginary field.\vspace{0.5in}

\textbf{REFERENCES}\vspace{0.3in}

[1] R. Penrose, \emph{The Emperor's New Mind: Concerning Computers, Minds,
and the Laws of Physics}, Oxford Univ. Press, 1989;-- \emph{Shadows of the
Mind: A Search for Missing Science of Consciousness}, Oxford Univ. Press,
1994

[2] F. Crick, \emph{The Astonishing Hypothesis: The Scientific Search for
the Soul}, Charles Scribner's Sons, 1994

[3] G. M. Edelman and G. Tononi, \emph{A Universe of Consciousness: How
Matter Becomes Imagination}, Basic Books, 2000

[4] E. Novikov, A. Novikov, D. Shannahoff-Khalsa, B. Schwartz, and J.
Wright, Phys. Rev. E, \textbf{56} (3), R2387 (1997); --Appl. Nonl. Dyn. and
Stoch. Systems (ed. J. Kadtke and A. Bulsara), p.299, Amer. Inst. Phys., N.
Y., 1997

[5] C. V. Johnson, \emph{D-Branes}, Cambridge Univ. Press, 2003

[6] E. A. Novikov, ``Energy from hidden time'', International Journal of
Paraphysics, \textbf{22} (1\&2), 19 (1988)

[7] E. A. Novikov, Phys. of Fluids, \textbf{15 }(9), L65 (2003)

[8] V. K. Jirsa, K. J. Jantzen, A. Fuchs, and J. A. S. Kelso, IEEE Trans. on
Medical Imaging, \textbf{21}(5), 497 (2002)

[9] Y. Fujii and K. Maeda, \emph{The Scalar-Tensor Theory of Gravitation},
Cambridge Univ. Press, 2003

\end{document}